\documentclass[11pt,reqno,a4paper]{amsart} 

\usepackage{amsmath, amstext, amssymb}
\usepackage[T1]{fontenc}
\usepackage[latin1]{inputenc}
\usepackage{multirow} 
\usepackage{subfig}

\usepackage{hyperref}

\usepackage{enumerate}
\usepackage{url}

\usepackage{color}
\usepackage{array}
\definecolor{shg}{rgb}{0.55,0.55,0.55}
\definecolor{hg}{rgb}{0.65,0.65,0.65}
\definecolor{mg}{rgb}{0.75,0.75,0.75}
\definecolor{dg}{rgb}{0.85,0.85,0.85}
\definecolor{wh}{rgb}{1,1,1}
\definecolor{sdg}{rgb}{0.95,0.95,0.95}

\def\hbm{\mbox{}\hfill$\boxdot$}

\def\Red{\textcolor{red}}
\definecolor{dcyan}{rgb}{0.22,0.53,0.47}
\definecolor{dblue}{rgb}{0.0,0.0,0.68}

	\def\nn{\nonumber}

\newcommand{\eq}[3]{\begin{equation}\label{#1}%
  \renewcommand{\arraystretch}{1.4}%
  \begin{array}{#2}#3\end{array}\end{equation}}

\usepackage{graphicx}
\usepackage{psfrag}
\usepackage{chemarrow}
\usepackage{rotating}
\usepackage{lscape}

\def\R{\ensuremath{I\!\!R}}
\def\Rnn{\ensuremath{I\!\!R_{\geq 0}}}
\def\Rp{\ensuremath{\R_{> 0}}}

\newtheorem{definition}{Definition}[section]
\newtheorem{remark}[definition]{Remark}

\newtheorem{prop}[definition]{Proposition}

\newtheorem{fact}[definition]{Fact}

\DeclareMathOperator{\sign}{sign}
\DeclareMathOperator{\im}{im}
\DeclareMathOperator{\col}{col}
\DeclareMathOperator{\diag}{diag}

\newcommand{\brac}[1]{\ensuremath{\left(#1\right)}}

\def\Y{{\mathcal{Y}}}

\newcommand{\Zi}[1]{\ensuremath{Z^{\brac{#1}}}}

\newcommand{\ei}[1]{\ensuremath{e^{\brac{#1}}}}

\newcommand{\e}[1]{\ensuremath{e^{\brac{#1}}}}

\newcommand{\Yo}[1]{\ensuremath{Y_0\brac{#1}}}

\renewcommand{\theequation}{\thesection.\arabic{equation}}

\begin{document}

\title{N-site phosphorylation systems with 2N-1 steady states}

\author{\mbox{Dietrich Flockerzi, Katharina Holstein and Carsten Conradi}}

\address{
  Max Planck Institute Dynamics of Complex Technical Systems, \
  Sandtorstrasse 1, \
  D-39106 Magdeburg,\
  Germany\\
}

\email{\{flockerzi,conradi\}@mpi-magdeburg.mpg.de, k\_holstein@posteo.de}

\begin{abstract}
  Multisite protein phosphorylation plays a prominent role in
  intracellular processes like signal transduction, cell-cycle control
  and nuclear signal integration. Many proteins are phosphorylated in
  a sequential and distributive way at more than one phosphorylation
  site. Mathematical models of $n$-site sequential distributive
  phosphorylation are therefore studied frequently.  In particular, in
  {\em Wang and Sontag, 2008,} it is shown that models of $n$-site
  sequential distributive phosphorylation admit at most $2n-1$ steady
  states. Wang and Sontag furthermore conjecture that for odd  $n$,
  there are at most $n$ and that, for even $n$, there are at most
  $n+1$ steady states. This, however, is not true: building on earlier
  work in {\em Holstein et.al., 2013}, we present a scalar determining
  equation for multistationarity which will lead to parameter values
  where a $3$-site system has $5$ steady states and parameter values
  where a $4$-site system has $7$ steady states. Our results therefore
  are counterexamples to the conjecture of Wang and Sontag. We
  furthermore study the inherent geometric properties of
  multistationarity in $n$-site sequential distributive
  phosphorylation: the complete vector of steady state ratios is
  determined by the steady state ratios of free enzymes and
  unphosphorylated protein and there exists a linear relationship
  between steady state ratios of phosphorylated protein.
   
\end{abstract}


\maketitle

\section{Introduction}
\label{sec:intro}

\renewcommand{\theequation}{\thesection.\arabic{equation}}
\setcounter{equation}{0}

\begin{figure}[!hb] 
  \centering
  \includegraphics[width=.9\textwidth]{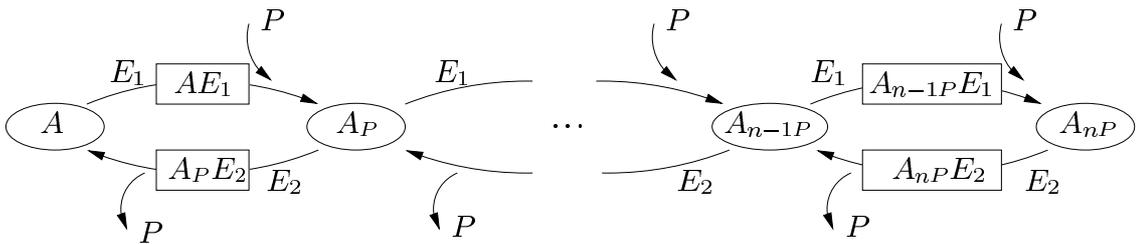}
  \caption{
    Sequential distributive phosphorylation and dephosphorylation of
    $A$ at $n$-sites by kinase $E_1$ and phosphatase $E_2$. Subscript
    $iP$ with $0\leq i \leq n$ denotes the phosphorylated forms of $A$
    (\lq phosphoforms\rq) and the number of phosphorylated sites (with
    $A = A_{0P}$ and $A_{p} = A_{1P}$). Each encounter of $A_{iP}$ and
    $E_1$ ($A_{iP}$ and $E_2$) results in at most one phosphorylation
    (dephosphorylation). Hence $n$ encounters of $A_{iP}$ and $E_1$
    ($A_{iP}$ and $E_2$) are required for phosphorylation
    (dephosphorylation) of $n$-sites. For biochemical details see, for
    example, \cite{ptm-003,sig-034,sig-041,sig-010}.
   }
  \label{fig-1}
\end{figure}

Protein phosphorylation and dephosphorylation are important
intracellular processes and many proteins are phosphorylated at more
than one phosphorylation site. Phosphorylation can either be
processive or distributive and sequential or random (see, for example,
\cite{ptm-003,ptm-025,sig-034,sig-041,sig-010}). Here we
focus on sequential distributive phosphorylation of 
a generic protein $A$ at $n$ sites by a kinase $E_1$ and its
sequential distributive dephosphorylation by a phosphatase $E_2$ (cf.\
Fig.~\ref{fig-1}). This process plays an important role in
signal transduction, cell-cycle control or nuclear signal integration
\cite{sig-034,sig-041}. A common interpretation of different (stable) 
steady states is that of an intracellular mechanism for information
storage \cite{bif-006,bif-013,ptm-005}. From this point of view, the
maximal possible number of steady states is an important quantity to
asses the information storage capacity of the system.

Under the assumption of  mass-action kinetics one obtains a polynomial
dynamical system in a straightforward way \cite{fein-043}. This
dynamical system consists of $3n+3$ ordinary differential equations
with polynomial right hand side involving $6n$ parameters. Its
variables represent the concentrations of the chemical species: kinase
$E_1$ and phosphatase $E_2$, unphosphorylated protein $A$ and the
phosphoforms $A_{iP}$, the kinase substrate complexes $A_{iP}E_1$ and
the phosphatase substrate complexes $A_{iP}E_2$. Of these $3n+3$
variables only $n+3$ can be measured with reasonable effort: the
concentration of $E_1$, $E_2$, $A$ and the $A_{iP}$. Hence parameter
values are subject to high uncertainty and one is either lead to apply
reductionist modeling approaches tailored to the system and question
at hand (as suggested, for example, in \cite{ptm-024}) or to studying
the whole parametrized family of polynomial ODEs (as, for example, in
\cite{BMB2013,ToricRN,multi-001} and the present publication).

The steady states of this parametrized family have been studied in a
variety of publications: Reference \cite{ptm-003} establishes a functional
relationship between the steady state ratio of kinase and phosphatase
on the one hand and the steady state value of the fully phosphorylated
protein on the other hand. The authors furthermore study the effect of
the number  $n$of phosphorylation sites on the graph of that
function. For fixed parameter values, the steady state values of
the phosphoforms $A_{iP}$ satisfy the algebraic relationships
described in \cite{ptm-001,ptm-002}. In particular, measurements of
the $A_{iP}$ taken from a given system (protein -- kinase --
phosphatase) have to satisfy these algebraic relations, provided the
system is distributive. These algebraic relations are therefore called
invariants in \cite{ptm-001,ptm-002}, and it is suggested to exploit
these invariants to discriminate different phosphorylation
mechanisms. In \cite{fein-067}, it is explained how such invariants can
be obtained for arbitrary biochemical reaction networks. The steady
states of post-translational modification systems, like the one
depicted in Fig.~\ref{fig-1}, admit a rational parameterization
\cite{ptm-006}. In \cite{ToricRN}, this has been specialized to the
system studied here:  it belongs to the class of chemical reaction
systems with toric steady states (defined in \cite{ToricRN}) and a
particular rational  parameterization is described. It is also shown
that, for such systems with toric steady states, necessary and
sufficient conditions for multistationarity (i.e.\ the existence of
multiple steady states) take the form of linear inequality systems.

The number of steady states has been studied in a variety of
publications as well. We start with results concerning $n=2$: here
bistability has been reported numerically for the first time in
\cite{sig-005}, multistationarity has been confirmed algebraically in
\cite{fein-012}. And in \cite{fein-018} it has been shown that 
multistationarity prevails in the presence of synthesis and
degradation of either kinase or phosphatase but not of both. An
implicit description of the region in parameter space where
multistationarity occurs is given in \cite{fein-024} and explicit
parameter conditions guaranteeing existence of three positive steady
states have been presented in \cite{maya-bistab}. For arbitrary $n$,
bistability has been established numerically in \cite{sig-034,sig-041}
and both, multistationarity and multistability have been reported in
\cite{sig-010}. The obvious fact that all phosphorylation sites
compete for the same kinase (phosphatase) has been described as a
possible explanation for the occurrence of multistationarity,
especially as the system depicted in Fig.~\ref{fig-1} lacks explicit
feedback loops; see \cite{fein-065} where this phenomenon is called
enzyme-sharing. Finally, in \cite{multi-001} it has been shown that
this system has at most $2n-1$ positive steady states. There the
authors also show the existence of parameter values where the system has
$n$ ($n+1$) steady states for $n$ even (odd) and conjecture that $n$
($n+1$) is an upper bound for the number of steady states. 
If, as described above, steady states are considered as an 
intracellular means to store information, then this conjecture asserts
that the achievable capacity of the system ($n$ or $n+1$ steady states resp.)
is far from the theoretical upper bound ($2n-1$). Later on, in
Section~\ref{sec:counterex}, we will provide counterexamples for $n=3$
and $n=4$. Hence the conjecture is not true in general, however, we do
not provide any information as to whether the theoretical maximum can
be achieved for biochemically meaningful parameter values.

In the previous publication \cite{BMB2013}, we have analyzed
multistationarity for arbitrary $n\geq 2$: there we present a collection of
feasible linear inequality systems and show that solutions of these
systems define parameter values where multistationarity occurs
(together with two steady states as witness). In the present
contribution, we combine the results of \cite{BMB2013} with ideas and
methods of proof from \cite{multi-001} to obtain in eq.~(\ref{polyp})
a 
univariate polynomial $P$ of degree $2n+1$ whose 
{\em admissible} positive zeros
are in one-to-one correspondence with positive steady
states.
Here, a positive zero $\xi_0$ of $P$ is called {\em admissible}
if a certain polynomial $G$ of degree $n$ is positive at $\xi_0$ 
(cf. Fact~\ref{admiss1}).
Multistationarity then requires $\geq 2$ admissible positive roots
of $P$.  By applying an argument already used in \cite{multi-001}
we can show that $P$ has at most $2n-1$ positive roots
(cf. Remark~\ref{rem:excase}).
 
Incorporating the admissibility condition, we pass from $P=0$ to a
scalar determining eqution $\theta = 0$ in
Proposition~\ref{prop:poly_sol} so that positive zeros of $\theta$ are
automatically admissible and thus in one-to-one correspondence with
positive steady states. For $n=3$ and $n=4$ we furthermore exploit the
structure of $\theta$ to explicitly construct parameter values where
$\theta$ has $5$ and $7$ positive roots (cf.\ Fig.~\ref{pitch_5ss} \&
\ref{nov28-7defs.1} and Table~\ref{numtab}). We also explain how the
same construction can be applied to obtain parameter values for at
least $n+1$ steady states for $n > 4$.

We further investigate the geometry of multistationarity: if
parameters are such that $\theta$ admits $\geq 2$ positive roots, then 
measurement of two different steady state values of kinase,
phosphatase and protein alone suffices to reconstruct the complete
vector of ratios of both steady states
(Fact~\ref{fact:computing_Gamma}).  We use this fact to devise a
graphical test based on measurement data to discard the possibility
that the measured data give rise to multistationarity
(Fact~\ref{fact:collinear} and Remark~\ref{rem:graph_excl}). In the
spirit of \cite{ptm-001,ptm-002} our results
Fact~\ref{fact:computing_Gamma} and \ref{fact:collinear} can be
interpreted as invariants characterizing steady states when parameter
values are in the multistationarity regime (as opposed to the
invariants described in \cite{ptm-001,ptm-002} that hold regardless of
whether or not parameters are in the multistationarity regime). To the
best of our knowledge these invariants have not been described before.
 
This paper is organized as follows: Section ~\ref{sec:notation} and
Section~\ref{sec:network} introduce the necessary notations and  the
basic facts from \cite{BMB2013}.  In the spirit of \cite{multi-001},
Section~\ref{sec:reduct} presents a scalar determining equation for
multistationarity which will be studied, in
Section~\ref{sec:counterex}, for an explicit triple phosphorylation
network possessing $5=2\cdot 3-1$ positive steady states. We also
present a $4$-site phosphorylation network with $7=2\cdot 4-1$
positive steady states. The concluding Section~\ref{sec:discuss}
discusses the geometry of multistationarity, addresses the
constraints on corresponding steady state ratios and comments on
measurement and reconstruction issues. In  Appendix \ref{sec:smat}, we
present explicit formulae for the network matrices associated to a
triple phosphorylation in Fig.~\ref{fig-1}.

\vspace{5mm}\section{\ Notation}
\label{sec:notation}
\renewcommand{\theequation}{\thesection.\arabic{equation}}\setcounter{equation}{0}

We use the symbol $\R^m$ to denote Euclidean $m$-space, the symbol
$\Rnn^m$ to denote the nonnegative orthant and $\Rp^m$ to denote the
interior of the nonnegative orthant. Vectors are considered as column
vectors and, for convenience, usually displayed as row vectors using
$^T$ to denote the transpose. For example,
$x\in \R^m$ will usually be displayed as $\brac{x_1,\, \ldots,\,    x_m}^T$.
The vector $x\in\R^m$ with $x_i=1$ for $i=1,...,m$ will be denoted by $\underline{1}$.

We will use the
  symbol $e_j$ to denote elements of the standard basis of Euclidian
  vector spaces and use the superscript $^{\brac{i}}$ to distinguish
  basis vectors of vector spaces of different dimension $3i+3$:
  \begin{align}
    \nn 
    \ei{i}_j &\text{\ldots denotes elements of the standard basis of
      $\R^{3i+3}$.} 
  \end{align}

For positive vectors $x\in\Rp^m$ we use the shorthand notation $\ln x$
to denote
  \begin{align}\nn 
    \ln x &:= \brac{\ln x_1,\, \ldots,\, \ln x_m}^T\, \in \, \R^m. \\\nn
    \intertext {Similarly, for $x\in\R^m$, we use $e^x$ to denote}\nn
    e^x &:= \brac{e^{x_1},\, \ldots,\, e^{x_m}}^T\, \in \, \Rp^m
    \intertext{and, for $x\in\R^m$ with $x_i\neq 0$, $i=1$, $\ldots$, $m$,}\nn
    x^{-1} &:= \brac{\frac{1}{x_1},\, \ldots,\, \frac{1}{x_m}}^T\, \in \, \R^m\, .
  \end{align}
Finally, $x^y$ with $x$, $y\in\Rnn^m$ will be defined by
\begin{equation}\nn 
  x^{y} := \prod_{i=1}^m x_i^{y_i}\, \in \, \Rnn\, .
\end{equation}

\vspace{5mm}\section{\ Steady states of a dynamical system
    derived from Figure~\ref{fig-1}}
\label{sec:network}
\renewcommand{\theequation}{\thesection.\arabic{equation}}\setcounter{equation}{0}

By describing every reaction at the mass action level, we derive a
dynamical system form Fig.~\ref{fig-1}. For this purpose we use the
notation introduced in \cite{BMB2013}. We also summarize those results of
\cite{BMB2013} that are relevant for this contribution. We would like to
emphasize that the dynamical system determined here and the one
considered in \cite{multi-001} are identical (up to a change of
variables).

The mass action network derived from Fig.~\ref{fig-1} (with $n$ an
arbitrary but fixed positive number) 
consists of the following $3+3n$ chemical species: the protein
(substrate) $A$ together with $n$ phosphoforms $A_P$, \ldots,
$A_{nP}$; the kinase $E_1$ together with $n$ kinase-substrate
complexes $A\, E_1$, \ldots, $A_{n-1P}\, E_1$ and the phosphatase
$E_2$ together with $n$ phosphatase-substrate complexes $A_{P}\, E_2$,
\ldots, $A_{nP}\, E_2$. 
To
each species, a variable $x_i$ denoting its concentration is
assigned:
\begin{eqnarray}
  \label{tab:VarAssignment}
  x_1=E_1\, ,\ x_2=A\, ,\ x_3=E_2\, , &x_{1+3i}=A_{(i-1)P}E_1\, ,\
  x_{2+3i}=A_{iP}\, ,\ x_{3+3i}=A_{iP}E_2 
\end{eqnarray}
with $A_{0P} = A$ and $A_{1P} = A_P$ ($i=1,...,n)$.
We collect all variables in a $(3+3n)$-dimensional vector
$x := \brac{x_1,\, \ldots,\, x_{3+3n}}^T$. As it will turn out, the
chosen labeling entails a simple block structure for the matrices
associated to the  dynamical system \eqref{eq:ode_def} of the network
in Fig.~\ref{fig-1}, cf., for example, the block structure
\eqref{eq:Ei} for the generators of the nonnegative cone in the kernel
of the stoichiometric matrix.

Assuming a distributive mechanism, a single phosphorylation occurs
with each encounter of substrate and kinase, and $n$ phosphorylations
therefore require $n$ encounters of substrate and kinase.
Similarly, $n$ dephosphorylations following a distributive mechanism
require $n$ encounters of substrate and phosphatase. Each
phosphorylation and each dephosphorylation therefore consists of 
3 reactions and consequently the network consists of $6n$
reactions. To each reaction we associate a rate constant. We use $k_i$
for phosphorylation  and $l_i$ for dephosphorylation reactions and 
obtain the following reaction network:
\begin{equation}
  \label{eq:network_dd}
  \begin{split}
    E_1 + A_{i-1P} \autorightleftharpoons{$k_{3i-2}$}{$k_{3i-1}$}
    A_{i-1 P}\, E_1 \autorightarrow{$k_{3i}$}{} E_1 + A_{iP},\quad 
    i=1,\, \ldots,\, n \\ 
    E_2 + A_{iP} \autorightleftharpoons{$l_{3i-2}$}{$l_{3i-1}$} A_{i
      P}\, E_2 \autorightarrow{$l_{3i}$}{} E_2 + A_{i-1P},\quad i=1,\,
    \ldots,\, n\ .
  \end{split}
\end{equation}
Using this notation, $k_{3i-2}$ ($l_{3i-2}$) denotes the association
constant, $k_{3i-1}$ ($l_{3i-1}$) the dissociation constant and
$k_{3i}$ ($l_{3i}$) the catalytic constant of the $i$-th
phosphorylation (dephosphorylation) step. We collect all rate
constants in a vector 
\begin{equation}\label{eq:def_kappa}
  \kappa := \col\brac{\kappa_{\brac{1}},\, \ldots,\,
    \kappa_{\brac{n}}}\in \Rp^{6n}
\end{equation}	
with the sub-vectors  $\kappa_{\brac{i}} := \brac{k_{3i-2},\,
  k_{3i-1},\, k_{3i},\,l_{3i-2},\, l_{3i-1},\, l_{3i}}^T$.

\vspace{3mm}
For every $n$, one can derive
the stoichiometric matrix $S\in \R^{\brac{3+3n} \times 6n}$ and the
rate exponent matrix \mbox{$\Y\in\R^{(3+3n)\times 6n}$} from
(\ref{eq:network_dd}), cf. \cite{BMB2013} for example. These define
two monomial functions and a dynamical system in the following way
where we denote the columns of $\Y$ with $y_i$: 
\begin{itemize}
\item Monomial functions $ \Phi:\R^{3+3n} \to 
  \R^{6n}$ and $r(\kappa,\cdot):\R^{3+3n}\to\R^{6n}$:
  \begin{equation}
    \label{eq:def_phinx}
    \Phi\brac{x}:=  x^{\Y^T}\equiv \brac{x^{y_1},\, \ldots,\,
      x^{y_{6n}}}^T 
    \quad \mbox{and} \quad
    r(\kappa,x) := \diag\brac{\kappa}\, \Phi\brac{x}\, . 
  \end{equation}
  The $6n$-dimensional vector $r(\kappa,x)$ is called the reaction
  rate vector.
\item Dynamical system:
  \begin{equation}
    \label{eq:ode_def}
    \dot x = S\, r(\kappa,x)\, =S\, \diag{(\kappa)}\, x^{\Y^T}\, .
  \end{equation}
\end{itemize}
If the three rows of a matrix $Z\in \R^{3\times (3+3n)}$ form a basis for the left kernel of $S$
-- as the three rows of the matrix $\Zi{n}$ defined in formula (9) of \cite{BMB2013} --
then the level sets 
$$\{x\in\R^{3+3n} :\, Z\, x = {const.}\}$$ 
are invariant under the flow of (\ref{eq:ode_def})
as one has $Z\, x(t) = Z\, x(0)$ along solutions $x(t)$ of
(\ref{eq:ode_def}). This observation motivates the classical
definition of multistationarity.

\vspace*{5mm}
\begin{definition}[Multistationarity]\label{defimulti} \mbox{}\\
  The system $\dot x = S\, r(\kappa,x)$ from \eqref{eq:ode_def} is
  said to exhibit multistationarity if and only if there exist a
  positive vector $\kappa \in \Rp^{6n}$ and at least two distinct positive
  vectors $a$, $b \in \Rp^{3+3n}$ with
  \begin{subequations}
    \begin{align}
      \label{eq:multistat_ode_x0}
      S\, r(\kappa,a) &= 0 \, ,\\
      \label{eq:multistat_ode_x1}
      S\, r(\kappa,b) &= 0 \, ,\\
      \label{eq:multistat_con_rel_x0_x1}
      Z\, a &= Z\, b.
    \end{align}
  \end{subequations}
\end{definition}
The equations \eqref{eq:multistat_ode_x0} and \eqref{eq:multistat_ode_x1}
describe the steady state property of $a$ and $b$ whereas
the equation \eqref{eq:multistat_con_rel_x0_x1} asks for these steady states 
to belong to the same coset of the stoichiometric matrix $S$.

\vspace{3mm}
For the purpose of this contribution, the monomial function
$\Phi$ and the matrix $Z$ are of particular interest. 
We refer to
Appendix~\ref{sec:smat} 
for
expressions defining the matrix $S$ and 
for the explicit model of network \eqref{fig-1}
for $n=3$ (cf.\ \cite{BMB2013}). 
Using the ordering of species and reactions introduced above in
equation~(\ref{tab:VarAssignment}) 
one obtains 
the matrix 
$Z\in \R^{3\times\brac{3+3n}}$ of conservation laws
and the rate exponent matrix $\Y\in \R^{\brac{3+3n} \times 6n}$
in the following way:

\vspace*{2mm}(I)  With
\begin{align}\nn
  \Yo{i} &:=\left[
    \begin{array}{cccccc}
      \e{i}_1 + \e{i}_{3i-1} & \e{i}_{3i+1} & \e{i}_{3i+1} &
      \e{i}_3 + \e{i}_{3i+2} & \e{i}_{3i+3} & \e{i}_{3i+3}
    \end{array}
  \right] \, \in \R^{(3+3i)\times 6}\, ,\end{align}
the rate exponent matrix $\Y\in\R^{(3+3n)\times 6n}$ is given by
\begin{equation}\label{eq:Y_direct}
  \Y^T:= \
  \left[
    \begin{array}{c|c|c|c|c|c|c}
      \Yo{1}^T & 0_{6\cdot 1 \times3} & \multirow{2}*{$0_{6\cdot
          2 \times 3}$} & \multirow{3}*{$0_{6\cdot 3 \times 3} $}&
      \phantom{0_{6\cdot 2 \times 3}} & \phantom{0_{6\cdot 2
          \times 3}} &  \multirow{5}*{$0_{6\cdot n \times 3}$} \\
      \cline{1-2}
      \multicolumn{2}{c|}{\Yo{2}^T} & & & & & \\ \cline{1-3}
      \multicolumn{3}{c|}{\Yo{3}^T} & & & \\
      \cline{1-4} \multicolumn{5}{r|}{\ddots} &  \\ \cline{1-5}
      \multicolumn{6}{c|}{\Yo{n-1}^T} &  \\ \hline
      \multicolumn{7}{c}{\Yo{n}^T}\\
    \end{array}
  \right]\, .
\end{equation}
 
\vspace*{2mm}(II)  
The matrix $Z\in \R^{3\times (3+3n)}$
of conservation laws is given by\\[1mm] 
\begin{equation}
  \label{eq:def_Zi}
  Z = \left[
  \begin{array}{cc}
      \begin{array}{rrr|}
        1 & 0 & 0\phantom{.} \\
        -1 &\phantom{-}1 & -1\phantom{.}\\
        0 & 0 & 1\phantom{.}
      \end{array}\,
    &
        \begin{array}{rrr|r|rrr}
        1 & 0 & 0\phantom{.}  & & 1 & 0 & 0 \\
        \phantom{.}0 & \phantom{.}1 & \phantom{.}0\phantom{.} & \phantom{.}\cdots\phantom{.} & \phantom{.}0 & \phantom{.}1 &\phantom{.} 0 \\
        0 & 0 & 1\phantom{.}  & & 0 & 0 & 1
        \end{array}
   \end{array}\right]  . 
  \end{equation}
We note that the three rows of the present $Z$ form a basis for the left kernel of $S$
as the three rows of the matrix $\Zi{n}$ defined in formula (9) of
\cite{BMB2013}. The first row of $Z$, for example, refers to the
conservation of the total $E_1$-concentration.

\vspace{3mm}
We now recall the discussion of the pointed polyhedral cone 
$\ker\brac{S} \cap {\Rnn^{6n}}$ (cf. Lemma 3.5 of \cite{BMB2013}) and
the computation of steady states (cf. Theorem 4.2 and Remark 4.3 of
\cite{BMB2013}). 
First, we define the  matrix 
\begin{equation}\label{eq:Ei}
  E := \left[
    \begin{array}{ccc}
      E_0 & & \\
      & \ddots & \\
      & & E_0
    \end{array}
  \right]\, \in \Rp^{6n\times 3n} \quad \mbox{with } \ E_0 := \left[
    \begin{array}{ccc}
      1&0&1\\
      1&0&0\\
      0&0&1\\ 
      0&1&1\\
      0&1&0\\
      0&0&1
    \end{array}
  \right] \
\end{equation}
so that the columns of $E$ form a basis of $\ker\brac{S}$. In addition,
the columns of $E$ are generators of $\ker\brac{S} \cap
\Rnn^{6n}$.
Secondly, we define the matrix     
  \begin{equation}
    \label{eq:def_Ln}
    L := \left[
      \begin{array}{c}
        L\brac{0} \\
        L\brac{1} \\
        \vdots \\
        L\brac{n}
      \end{array}
    \right]\, \in \mathbb{Z}^{\brac{3+3n}\times 3}
    \ \mbox{ for} \   
    L\brac{0} :=
    \begin{bmatrix}
      \phantom{-}1 &\ n-1 &\ -1\\
      -1 &\ -n &\ \phantom{-}0\\
      \phantom{-}1 &\ n-2 &\ -1
    \end{bmatrix}\, , \ 
    L\brac{i} :=
    \begin{bmatrix}
      \phantom{-}0 &\ i-2 &\ -1 \\
      -1 &\ i-n & \ \phantom{-}0\\
      \phantom{-}0 &\ i-2 &\ -1
    \end{bmatrix} 
  \end{equation}

and observe that the matrix $L$ has the same range as the matrix 
$M$ defined in \cite[eqns.~(17a)--(17c)]{BMB2013} because of
\begin{equation}\nn 
  M=LR \quad \mbox{with } \ R=\begin{bmatrix}
    -1           & \phantom{-}0 & \phantom{-}0\\
    \phantom{-}0 & -1           & \phantom{-}1\\
    \phantom{-}0 & \phantom{-}0 & -1
  \end{bmatrix}\, .
\end{equation}

This choice of $L$ will turn out to be advantageous in the next section
since all entries of the first and third column come from $\{-1$, $0$,
$1\}$. Now we can summarize those points of \cite{BMB2013} 
that are relevant for the following discussion:

\begin{prop}[Multistationarity]\label{prop:multig}\mbox{}\rm \\
Recalling the dynamical system
\eqref{eq:ode_def}
and 
the matrices $Z$, $E$  and $L$ from (\ref{eq:def_Zi}),  (\ref{eq:Ei})  and \eqref{eq:def_Ln}
one has the following equivalences:
\begin{enumerate}
\item
A given $a\in\Rp^{3+3n}$ is a positive steady state of $\dot{x}=S\, r(\kappa,x)$  
if and only if there exists a $\lambda \in\Rp^{3n}$  with
\begin{equation}\label{k}
\kappa \ = \  \kappa(a,\lambda):= \diag\brac{a^{-\Y^T}}\, E\, \lambda\, .
\end{equation}
\item
A given $b\in\Rp^{3+3n}$ is a positive steady state of $\dot{x}=S\, r(\kappa(a,\lambda), x)$
if and only if
\begin{equation}\label{eq:def_b}
\ln\brac{b} \, - \, \ln\brac{a}\ \in \im\brac{L}
\end{equation}
holds true, i.e., if and only if there exists a $g\in \Rp^3$ with
\begin{equation}\label{eq:def_bg}
b\, =\, \diag(g^L)\, a\, .
\end{equation}
\item
Two positive steady states $a$ and $b=\diag(g^L)\, a$, $g\in\Rp^3$, of
$\dot{x}=S\, r(\kappa(a,\lambda), x)$ satisfy $Za=Zb$ from
\eqref{eq:multistat_con_rel_x0_x1} {\em if and only if}
$g\in \Rp^3$ is a solution of the 3-dimensional {\em coset condition} 
\begin{equation}\label{coset0}
    \Theta(g,a):=Z\, \big(\diag(g^L)-I\big)\, a\, =\,
    Z\, \diag(a)\, \big(g^L-\underline{1}\big)\, =\, 0\,, \quad
    g\in \Rp^3\, .
  \end{equation} 
   For $g\neq\underline{1}$, the steady states $a$ and 
  $b:=\diag(g^L)\,a$ 
  are distinct positive steady states for the 
  network $\dot{x}=S\, r(\kappa(a,\lambda), x)$
  within the same coset of the stoichiometric matrix $S$.
\end{enumerate}\end{prop}

The proof follows directly from \cite{BMB2013}.
For part (2), we just note that the existence of a $\mu \in \im\brac{L}$ with 
$\ln\brac{b} \, - \, \ln\brac{a}=\mu$, i.e., $b=\diag\brac{e^\mu}a$,
can be formulated with
\begin{equation}\label{varg1}
  \mu=L\, \ln \brac{g}\quad \mbox{for } \ 
  g=(g_1,g_2,g_3)^T\in \Rp^3
\end{equation} 
as \eqref{eq:def_bg} because of
\begin{equation}\nn
  g^L=(g^{L_1},...,g^{L_{3+3n}})^T=(e^{\ln\brac{g}})^L=e^\mu\, .
\end{equation}

We'd like to point out that the matrix $L$ in (\ref{eq:def_Ln})
constrains the components of $g^L$ and thus imposes a special geometry
on the steady states $a$  and $b$. 
For a biological interpretation, we refer to the discussion in Section~\ref{sec:discuss}.%


\vspace{5mm}\section{\ A scalar determining equation for multistationarity}\label{sec:reduct}
\renewcommand{\theequation}{\thesection.\arabic{equation}}\setcounter{equation}{0}

The previous section shows that multistationarity for the                                 
system \eqref{eq:ode_def}, derived from network~(\ref{eq:network_dd}),
can be characterized by the 3-dimensional coset condition
\eqref{coset0}. In the spirit of
\cite{multi-001}, we will prove that the simple form \eqref{eq:def_Zi} 
of the matrix $Z$, representing the conservation laws, allows  a
reduction to a scalar  equation 
\begin{equation}\nn 
  P(\xi,a)=0 
\end{equation}
where $P(\xi,a)$ is a polynomial in $\xi:=g_2$, the second component of $g$ (cf.  
the representations 
\eqref{polyphsum} and \eqref{polypabc} below).
A zero $\xi_0=\xi_0(a)$ of $P$ will be called an {\em admissible} zero (for
\eqref{coset0}) if and only if the corresponding $g=g(\xi_0(a),a)$
belongs to $\Rp^3$, i.e., if and only if the zero $\xi_0=\xi_0(a)$  is
positive and a certain scalar polynomial inequality $G(\xi_0(a),a)>0$
holds true (see Fact~\ref{admiss1} and \eqref{genps5G} below).

\vspace{2mm}
We first turn to the matrix $L$ of equation~(\ref{eq:def_Ln}),
denote the second column of $L(i)$ by $\ell_{(i)}$
and define 
\begin{equation}\nn 
\ell=\big(\ell_1,...,\ell_{3+3n}\big)^T=\big(\ell^T_{(0)},...,\ell^T_{(n)})\big)^T\in \mathbb{Z}^{3+3n}
\end{equation}
with $\ell^T_{(0)}=(n-1,-n,n-2)$ and $\ell^T_{(i)}=(i-2,i-n,i-2)$ for $i=1,...,n$.
Moreover we introduce 
\begin{equation}\label{omegaa}
\omega\, =\, (\omega_1,\omega_2,\omega_3)^T\, :=\, Za
\end{equation}
with the total enzyme concentrations $\omega_1=\sum_{k=0}^n\, a_{1+3k}$ and
$\omega_3=\sum_{k=0}^n\, a_{3+3k}$. For a more compact notation,
we suppress the dependence on $a\in \Rp^{3+3n}$ for the moment.

The 3-dimensional system  \eqref{coset0} can thus be written as
\begin{subequations}\label{traf0}
\begin{eqnarray}\label{genps2}
\omega_1\ = & g_3^{-1}\Big[a_1g_1\xi^{\ell_1}+a_4\xi^{\ell_4}+\cdots+ a_{1+3n}\xi^{\ell_{1+3n}}\Big]
\,   ,\\\label{genps22}
\omega_3\ = & g_3^{-1}\Big[a_3g_1\xi^{\ell_3}+a_6\xi^{\ell_6}+\cdots+ a_{3+3n}\xi^{\ell_{3+3n}}\Big]\phantom{\, ,}
\end{eqnarray}
\end{subequations}
together with
\begin{eqnarray}\label{genps23}\omega_2= & 
-a_1g_1g_3^{-1}\xi^{\ell_{1}}-a_3g_1g_3^{-1}\xi^{\ell_{3}}
+g_1^{-1}\Big[a_2\xi^{\ell_{2}}+\cdots+ 
a_{2+3n}\xi^{\ell_{2+3n}}\Big]\, . 
\end{eqnarray}
Because of $\ell_1=1+\ell_3$, the system \eqref{traf0} can now be written as
\begin{subequations}\label{genps3}
\begin{eqnarray}\label{genps3a}
a_1\xi\cdot g_1\xi^{\ell_{3}}-\omega_1 g_3{=}&-\Big[
a_4\xi^{\ell_{4}}+\cdots+ 
a_{1+3n}\xi^{\ell_{1+3n}}\Big]\,   ,\\\label{genps3b}
a_3\cdot g_1\xi^{\ell_{3}}-\omega_3 g_3{=}&-\Big[
a_6\xi^{\ell_{6}}+ \cdots+
a_{3+3n}\xi^{\ell_{3+3n}}\Big]\,   .
\end{eqnarray}
\end{subequations}
  We exploit the structure of the subvectors $\ell_i$ to represent the
  system (\ref{genps3}) as a $\xi$-dependent linear
  system for  $g_1$, $g_3$. For this purpose we introduce the polynomials
  \begin{subequations}\label{omexi}
    \begin{equation}
      \label{omexi13}
      \Omega_4(\xi)=a_4+a_7\xi+\cdots +a_{1+3n}\xi^{n-1}\, , \
      \Omega_6(\xi)=a_6+a_9\xi+\cdots +a_{3+3n}\xi^{n-1}
    \end{equation}
    and note the  relations to 
    \begin{equation}\label{omexi13a}
      \omega_1=a_1+\Omega_4(1)\, ,\ \omega_3=a_3+\Omega_6(1).
    \end{equation}
    For later purposes, we also introduce the $n$-th order polynomial
    \begin{equation}
      \label{omexi2}
      \Omega_2(\xi):=a_2+ a_{5}\xi+ \cdots  +a_{2+3n}\xi^n
      \quad \mbox{with } \ \omega_2=-a_1-a_3+\Omega_2(1)\, ,
      \end{equation}
    where $\omega_2$ is not necessarily positive (cf.~\eqref{omegaa}). 
  \end{subequations}
  With the help of $\Omega_4\brac{\xi}$ and $\Omega_6(\xi)$ the system
  (\ref{genps3}) reads
  \begin{displaymath}
    \left[
      \begin{array}{cc}
        a_1\, \xi^{n-1} & -\omega_1 \\
        a_3\, \xi^{n-2} & -\omega_3
      \end{array}
    \right] 
    \left(
      \begin{array}{c}
        g_1 \\ g_3
      \end{array}
    \right) = -\frac{1}{\xi}\, 
    \left(
      \begin{array}{c}
        \Omega_4\brac{\xi} \\
        \Omega_6\brac{\xi}
      \end{array}
    \right).
  \end{displaymath}

If
\begin{equation}\label{polyxib}
  \Delta{(\xi)}:=\frac{a_1\xi}{\omega_1}-\frac{a_3}{\omega_3}=\frac{a_1}{\omega_1}(\xi-\xi^*)
\end{equation}
  is nonzero, that is, if
\begin{equation}\label{polyxibb}
  \xi\neq \xi^*=\frac{\omega_1/a_1}{\omega_3/a_3}\, >0\,  ,
\end{equation}
then system \eqref{genps3} possesses the unique solution
\begin{subequations}
  \label{genps4}
  \begin{eqnarray}\label{genps4a}
    g_1=g_1(\xi):=&\xi^{1-n}F_1(\xi)/\Delta(\xi)\, 
    , \\\label{genps4b}
    g_3=g_3(\xi):=&\xi^{-1}F_3(\xi)/\Delta(\xi)\ \
  \end{eqnarray}
  for the following polynomials $F_1$ and $F_3$ in $\xi$ of degree
  $n-1$ and  $n$, respectively:
\end{subequations}

\begin{subequations}
  \label{polyxif}
  \begin{eqnarray}
    \label{polyxic1}
    F_1(\xi):=&\dfrac{\Omega_6(\xi)}{\omega_3}-\dfrac{\Omega_4(\xi)}{\omega_1}
    \\\label{polyxic3}
    F_3(\xi):=&
    \dfrac{a_1\xi}{\omega_1}\,\dfrac{\Omega_6(\xi)}{\omega_3}-
    \dfrac{a_3}{\omega_3}\,\dfrac{\Omega_4(\xi)}{\omega_1} \, 
  \end{eqnarray}
\end{subequations}

Concerning polynomials $F_1$ and $F_3$ we observe the following identities
  \begin{equation}
    \label{eq:polyrel}
    F_3(\xi) \,=\,  
    \dfrac{\Omega_4(\xi)}{\omega_1}\,\Delta(\xi)+\dfrac{a_1\xi}{\omega_1}F_1(\xi) 
    \,=\,  
    \dfrac{\Omega_6(\xi)}{\omega_3}\,\Delta(\xi)+\dfrac{a_3}{\omega_3}F_1(\xi)\, .
  \end{equation}

We note that, in case of \eqref{polyxibb}, $g_1$ is positive for positive $\xi$ if and only if
$F_1(\xi)$ and $\Delta(\xi)$ are of the same sign. By \eqref{genps4b} and \eqref{eq:polyrel},
$g_3$ is positive for such positive $g_1$. 

\begin{fact}[Positivity of $(g_1, g_3)$]\label{admiss1}\mbox{}\\
Given a positive $\xi$ with   $\xi\neq \xi^*$,
the  \eqref{genps3}-solution $(g_1(\xi), g_3(\xi))$, given by \eqref{genps4}, is positive if and only if
  \begin{equation}\label{admiss1G}
    G(\xi):=F_1(\xi)\Delta(\xi)>0
  \end{equation}
  holds true. 
\end{fact} 
For the case $\xi=\xi^*$, we refer to Remark~\ref{rem:excase}(b) below.
In what follows, we assume  \eqref{polyxibb} to be true.

\vspace*{3mm}
If these rational solutions \eqref{genps4} of the linear system \eqref{genps3} 
are inserted into \eqref{genps23}
one arrives -- with the notations~\eqref{omexi} -- at the equivalent
$(2n+1)$--order polynomial equation 
\begin{eqnarray}\nn 
  Q(\xi):= \
  \Delta(\xi)\,F_3(\xi)\frac{\Omega_2(\xi)}{\omega_2}\, - \, 
  \Big[\frac{a_1\xi+a_3}{\omega_2}\,\xi\, \big[F_1(\xi)\big]^2 
  \, + \, 
  \xi\,F_1(\xi)\,F_3(\xi)\Big]   \stackrel{!}{=}\ 0\, .
\end{eqnarray}
By the $F_1$-representations of $F_3$ in \eqref{eq:polyrel}, 
$Q$ can be written as 
\begin{eqnarray}\nn 
  Q_1(\xi)= 
  \Delta^2(\xi)\dfrac{\Omega_4(\xi)}{\omega_1}\,\frac{\Omega_2(\xi)}{\omega_2} -  
  \xi\, \Big[\frac{a_1\xi+a_3}{\omega_2}+ \dfrac{a_1\xi}{\omega_1}\Big]F_1^2(\xi)  -
  \Delta(\xi)F_1(\xi)\,\Big[\frac{\Omega_2(\xi)}{\omega_2}\,\dfrac{a_1\xi}{\omega_1}+
  \xi\dfrac{\Omega_4(\xi)}{\omega_1}\Big]
\end{eqnarray}
and as
\begin{eqnarray}\nn
  Q_3(\xi)= 
\Delta^2(\xi)\dfrac{\Omega_6(\xi)}{\omega_3}\,\frac{\Omega_2(\xi)}{\omega_2}-  
\xi\, \Big[\frac{a_1\xi+a_3}{\omega_2}+ \dfrac{a_3}{\omega_3}\, \Big]F_1^2(\xi)\,  - 
 \Delta(\xi)F_1(\xi)\,\Big[\frac{\Omega_2(\xi)}{\omega_2}\,\dfrac{a_3}{\omega_3}+
\xi\dfrac{\Omega_6(\xi)}{\omega_3}\Big].
\end{eqnarray} 

\vspace*{3mm}
We now take a  linear combination of 
these expressions 
with nonnegative scalars $h_1$ and $h_3$, $h:=(h_1,h_3)\neq
(0,0)$,
and define
\begin{subequations}\label{polyphsum}
\begin{equation}\label{polyph}
P_h(\xi):=\omega_2h_1Q_1(\xi)+\omega_2h_3Q_3(\xi)\, =\,
A_h(\xi)\Delta^2(\xi)+B_h(\xi)\Delta(\xi)F_1(\xi)-C_h(\xi)F_1^2(\xi)
\end{equation}
for
\label{polyxia3sv}
\begin{eqnarray}\label{genps5dAv}
A_h(\xi)\, = &\Big(h_1\frac{\Omega_4(\xi)}{\omega_1}+h_3\frac{\Omega_6(\xi)}{\omega_3}\Big)\,\Omega_2(\xi)\, ,
\phantom{---------....}\\\label{genps5dBv}
B_h(\xi)\, = &\Big(h_1\frac{a_1\xi}{\omega_1}+h_3\frac{a_3}{\omega_3}\Big)\,\Omega_2(\xi)-
\Big(h_1\frac{\Omega_4(\xi)}{\omega_1}+h_3\frac{\Omega_6(\xi)}{\omega_3}\Big)\,\omega_2\xi \, ,\\\label{genps5dCv}
C_h(\xi)\, = &\xi\,\Big[(h_1+h_3)(a_1\xi+a_3)+\Big(h_1\frac{a_1\xi}{\omega_1}+h_3\frac{a_3}{\omega_3}\Big)\omega_2\Big].
\phantom{......}
\end{eqnarray} 
\end{subequations}
Since $L$ is a matrix with integer entries, \eqref{traf0} and \eqref{genps23}
make sense for all $g$ with $g_j\neq 0$, $j=1,2,3$. Hence $P_h$ can be considered as a
function of $\xi \in \R$. By Fact~\ref{admiss1}, a zero  $\xi_0$ of $P_h$ with
$\xi_0\neq \xi^*$
will be called an {\em admissible} zero (for
\eqref{coset0}) if 
\begin{equation}\label{genps5G}
\xi_0>0 \quad \mbox{and}\quad G(\xi_0)=F_1(\xi_0)\Delta(\xi_0)>0
\end{equation}
hold true. Obviously, $\xi=1$ is an admissible zero of $P_h$ with
$g_1(1)=1=g_3(1)$ in case of $\xi^*\neq 1$.

In the special 
case with $h_1=\omega_1$ and $h_3=\omega_3$ in~\eqref{polyxia3sv},  one has
\begin{subequations}\label{polypabc}\begin{equation}\label{polyp}
P(\xi):=\omega_1\omega_2Q_1(\xi)+\omega_2\omega_3Q_3(\xi)\, =\,
A(\xi)\Delta^2(\xi)+B(\xi)\Delta(\xi)F_1(\xi)-C(\xi)F_1^2(\xi)
\end{equation}
for the polynomials
\begin{eqnarray}\label{polypA}
A(\xi)\, := &\Big(\Omega_4(\xi)+\Omega_6(\xi)\Big)\,\Omega_2(\xi)\, ,
\phantom{-------.....}\\\label{polypB}
B(\xi)\, := &\Big(a_1\xi+a_3\Big)\,\Omega_2(\xi)-
\Big(\Omega_4(\xi)+\Omega_6(\xi)\Big)\,\omega_2\xi \, ,\\\label{polypC}
C(\xi)\, := &\xi\,\Big[a_1\xi+a_3\Big]\Big[\omega_1+\omega_2+\omega_3\Big]\phantom{-----.....}
\end{eqnarray} 
\end{subequations}
of degree $2n-1$, $n+1$ and $2$, respectively.
Concerning an upper bound for the number of admissible zeros of \eqref{polyp}, we
refer to Remark~\ref{rem:excase}.

\vspace*{3mm} We observe that $P(\xi)=0$ 
can be viewed as a quadratic equation for
$F_1(\xi)/\Delta(\xi)$,
i.e. for $\xi^{n-1}g_1(\xi)$ (cf.~\eqref{genps4a}).
As a consequence of Fact~\ref{admiss1} and the positivity of $A$ and $C$ on $\Rp$,
we arrive at the following {\em admissibility result} for
equation \eqref{polyp} and hence of the coset  condition
\eqref{coset0}. Positive solutions $g$ of  \eqref{coset0} are
characterized by the {\em scalar determining equation}
$\theta(\xi,a)=0$ in \eqref{genps5f1} whereby we explicitly mention
the dependence on $a\in \Rp^{3+3n}$:

\vspace*{5mm}
\begin{prop}[Determining equation for $\xi>0$, $\xi\neq \xi^*$]\label{prop:poly_sol}\mbox{} \\
The determining equation for admissible solutions $g\in \Rp^3$ of the
coset condition \eqref{coset0} is  given by
\begin{equation}\label{genps5f1}
  \theta(\xi,a):=2C(\xi,a)F_1(\xi,a){-}\  \Delta(\xi,a)
  \big[B(\xi,a)+\big(B^2(\xi,a)+4A(\xi,a)C(\xi,a)\big)^{1/2}\big]\, =
  \, 0
\end{equation}
for the polynomials $A$, $B$ and $C$ from~(\ref{polypabc}).
Any positive zero $\xi=\xi(a)$ of $\theta(\xi,a)$, different from $\xi^*(a)$,
defines a positive steady state
$$b=\diag\brac{g^L} a  \neq \ a$$ 
of the network \eqref{eq:ode_def}
for $g=(g_1(\xi(a),a),\xi(a),g_3(\xi(a),a))^T$ from \eqref{genps4}.
\end{prop}

\vspace*{5mm}
\begin{remark}[At most $2n-1$ admissible zeros (cf. \cite{multi-001})]\label{rem:excase}\mbox{}\rm
\begin{subequations}\label{remex1}
\begin{enumerate}\item[(a)]
We first assume \eqref{polyxibb}, i.e., $\xi\neq\frac{\omega_1/a_1}{\omega_3/a_3}$.
Since the leading coefficient of $P(\xi)$ is positive and $P(0)$ is positive,
there exists at least one negative zero $\xi_{-1}$ of $P$. Obviously, $P(\xi^*)$ is negative.
We suppose that $P(\xi)$ has $2n$ distinct positive zeros  and that $P$ is negative
on an interval $(\xi',\xi'')\ni\xi^*$ with $P(\xi')=0=P(\xi'')$, $\xi'>0$. In case $F_1(\xi)$
has a zero $\xi^\#\in (\xi',\xi'')$, the value 
$P(\xi^\#)=\Delta^2(\xi^\#)\Omega_6(\xi^\#)\Omega_2(\xi^\#)/\omega_3$ would be positive.  
Hence $F_1$ cannot change its sign on $(\xi',\xi'')$.
The $g_1$-expression \eqref{genps4a} thus implies that only one of the values 
$g_1(\xi')$ and $g_1(\xi'')$ is positive. Summarizing, 
$P(\xi)$ has at most $2n-1$ positive zeros
under \eqref{polyxibb} (cf. \cite{multi-001} where this kind of 
argument has been introduced).
\item[(b)]
We now turn to the case $\xi=\frac{\omega_1/a_1}{\omega_3/a_3}$ with $\Delta(\xi)=0$ (cf.~\eqref{polyxib}).
System \eqref{traf0} is solvable if and only if
\begin{equation}\label{solv0}
\omega_1/\omega_3=\Omega_4(\xi)/\Omega_6(\xi)\, .
\end{equation}
Under \eqref{solv0}, 
$F_1$ and $F_3$ vanish at $\xi$ (cf.~\eqref{polyxif}) and 
the positive solution of \eqref{traf0}  is of the form
\begin{equation}\label{solv1}
g_1=g_1(\eta):=\xi^{1-n}\eta \, ,\ 
g_3=g_3(\eta):=
\frac{1}{\xi}\,
\Big(\dfrac{\Omega_4(\xi)}{\omega_1}+ \dfrac{a_1\xi}{\omega_1}\, \eta \Big)\, =\,
\frac{1}{\xi}\,
\Big(\dfrac{\Omega_6(\xi)}{\omega_3}+ \dfrac{a_3}{\omega_3}\, \eta \Big)
\end{equation}
with $\eta >0$.
Equation \eqref{genps23} is thus equivalent to
\begin{equation}\label{solv2}
A(\xi)+B(\xi)\eta - C(\xi)\eta^2=0
\end{equation}
with positive $C(\xi)$ and positive $A(\xi)$, cf.~(\ref{polypabc}). Hence there exists a unique positive zero $\eta_0$
of \eqref{solv2}. Consequently, this value of $\xi$ can yield at most one positive solution
$g$ of \eqref{coset0}.
We note that, under \eqref{solv0}, this $\xi$ is a zero of $P$ of order at least $2$. 
\end{enumerate}
\end{subequations}
Hence we conclude that $P(\xi)$ has at most $2n-1$ admissible zeros. We might add, as a side remark,
that  $a=\alpha\cdot \underline{1}$, $\alpha>0$, is the unique positive steady state of \eqref{eq:ode_def}  since
\eqref{polyp} 
is equivalent to $(\xi^n-1)(\xi^{n+1}-1)=0$ possessing just $\xi=1=\xi^*$ as positive (double) zero.\hbm
\end{remark}

\vspace*{3mm}
Summarizing, by an argument similar to the one of 
\cite{multi-001} we have shown, that the $a$-dependent polynomial
$P(\xi)$ in \eqref{polyp} possesses at most $2n-1$ distinct admissible
zeros so that there are at most $2n-1$ distinct steady states of
\eqref{eq:ode_def} within one coset of the stoichiometric matrix
$S$. Moreover we have established that
the distinct zeros of $\theta(\xi,a)$ in \eqref{genps5f1}
give rise to distinct steady states of \eqref{eq:ode_def} 
within one coset. 
Finally, we note that the choices $h_1=0$ and $h_3=\omega_3$ in \eqref{polyph}
lead to an analogous result in case of
\begin{equation}\label{om2+3}
\omega_2(a)+\omega_3(a)=\sum_{k=0}^na_{2+3k}\,+\, \sum_{k=1}^na_{3+3k}\, - \, a_1\, >\, 0\, .
\end{equation}
For
\begin{subequations}\label{polyxia3s2}
\begin{eqnarray}\label{genps5dA2}
A_0(\xi)\, := &\Omega_6(\xi)\,\Omega_2(\xi)\, ,\phantom{--.....}\\\label{genps5dB2}
B_0(\xi)\, := &a_3\Omega_2(\xi)-\omega_2\xi\Omega_6(\xi)\, ,\\\label{genps5dC2}
C_0(\xi)\, := &\xi\,\Big[\big(a_1\xi+a_3\big)\omega_3+a_3\omega_2\Big]\ .
\end{eqnarray} 
the distinct zeros of 
\begin{equation}\label{genps5f1om}
\theta_0(\xi,a):=2C_0(\xi,a)F_1(\xi,a){-}\  
\Delta(\xi,a) \big[B_0(\xi,a)+\big(B_0^2(\xi,a)+4A_0(\xi,a)C_0(\xi,a)\big)^{1/2}\big]\, = \, 0\, 
\end{equation}
\end{subequations}
give rise to distinct steady states of \eqref{eq:ode_def} 
within one coset. To this end, we just observe
that $A_0(\xi)$ is positive for $\xi>0$ and that \eqref{om2+3} entails the positivity of $C_0(\xi)$
for $\xi>0$.
In the following section we apply the determining equation \eqref{genps5f1om} to construct a triple phosphorylation network with
more than $3$ steady states. Obviously, 
the choices $h_1=\omega_1$ and $h_3=0$ in \eqref{polyph} entail an analogous result.


\vspace{5mm}\section{\ Phosphorylation systems with the maximal number
  of steady states}\label{sec:counterex}
\renewcommand{\theequation}{\thesection.\arabic{equation}}\setcounter{equation}{0}

We consider phosphorylation systems with $n$ sites for $n=2,3$ and $4$ and give examples
of multistationarity with the maximal number $2n-1$ of steady
states. For $n=2$ we refer to the Example~4.8 in 
\cite{BMB2013}. We continue with the case $n=3$.
Suppressing the $a$-dependence, $\theta_0(\xi,a)=0$ from
\eqref{genps5f1om}  can be written as
\begin{eqnarray}\nn 
  2C_0(\xi)\big[\omega_1\Omega_6(\xi)-\omega_3\Omega_4(\xi)\big]\
  +&a_3\omega_1\phantom{\xi}\big[B_0(\xi)+\big(B_0^2(\xi)+4A_0(\xi)C_0(\xi)\big)^{1/2}\big] 
  \\ \nn
  {=} &a_1\omega_3\xi\big[B_0(\xi)+\big(B_0^2(\xi)+4A_0(\xi)C_0(\xi)\big)^{1/2}\big]
\end{eqnarray}
where the $n$ parameters $a_{3j+1}$, $j=1,2,...,n$, appear just on the left-hand side 
and in a {\em linear} way. So they might be tuned to fulfill some prescribed constraints.
This fact is the main motivation for passing from the polynomial
description \eqref{polypabc} to the determining equations
\eqref{genps5f1} or \eqref{genps5f1om}.

For the triple phosphorylation,
we choose a positive $a\in \Rp^{3\cdot 3+3}$ and  fix the rate constant vector 
\begin{equation}\nn 
\kappa=\kappa (a)=\diag\brac{a^{-\Y^T}}E\, \underline{1}
\end{equation}
so that $a$ is a positive steady state of the network \eqref{eq:ode_def}. Obviously, one has $\theta_0(1,a)=0$.
In particular, we choose $a$ of the form
\begin{equation}\label{n3exa1}
a^*=\big(1,1,1|a_4,1,1|a_7,1,0.1|a_{10},0.32,60\big)^T\in \Rp^{12}
\end{equation}
and compute {\em analytically} the remaining $n=3$ parameters $a_4$, $a_7$ and $a_{10}$ so that 
$\theta_0(\xi,a^*)$ has the triple zero  $\xi=1$
and a further (simple) zero $\xi=\frac{1}{2}$, i.e., so that $n=3$
constraints are met.
That is, we solve the equations
  \begin{displaymath}
    \left.\frac{\partial}{\partial \xi}\theta_0(\xi,a^*)\right|_{\xi=1} = 0\, ,\ \
    \left.\frac{\partial^2}{\partial \xi^2}\theta_0(\xi,a^*)\right|_{\xi=1} = 0\, ,\ \ 
    \theta_0(\frac{1}{2},a^*) = 0
  \end{displaymath}
  and obtain the analytical solution:
  \begin{align*}
    a_4 &= \frac{7787061638}{39861237827} - 
    \frac{10658368327 \left(1129320903987944456 - 14276293028087
        \sqrt{5505644539} \right)} 
    {79722475654 \left(-340903663256564611+4572282020317
        \sqrt{5505644539}\right)}\\
    a_7 &= -\frac{1129320903987944456-14276293028087
      \sqrt{5505644539}}
    {20 \left(-340903663256564611+4572282020317
        \sqrt{5505644539}\right)} \\
    a_{10} &= \frac{476228483659}{39861237827} - 
    \frac{221291854961 \left(1129320903987944456 - 14276293028087
        \sqrt{5505644539}\right)}
    {39861237827 \left(-340903663256564611+4572282020317
        \sqrt{5505644539}\right)}
  \end{align*}
The resulting numerical values (up to 4 decimals) are given by
\begin{equation}\label{n3exa2}
a_4:=a_4^*=5.9026(84)...\, , \ a_7:= a_7^*=2.1344(85)...\, , \ a_{10}:=a_{10}^*=248.9413(34)...\ .
\end{equation}
The inequality \eqref{om2+3} is obviously satisfied.
The numerical value of the rate constant vector $\kappa=\kappa(a^*)$ is 
$$(
2, 
0.1694.., 
0.1694.. 
|
2, 
1, 
1 
|
2, 
0.4684..,
0.4684.. 
|
2, 
10, 
10 
|
2, 
0.0040.., 
0.0040.. 
|
6.25, 
0.0166.., 
0.0166..)^T$$
and the numerical value of $\xi^*$ at $a^*$ is $4.1542....$.

\begin{figure}[h]
  \centering
  \includegraphics[width=.8\linewidth]{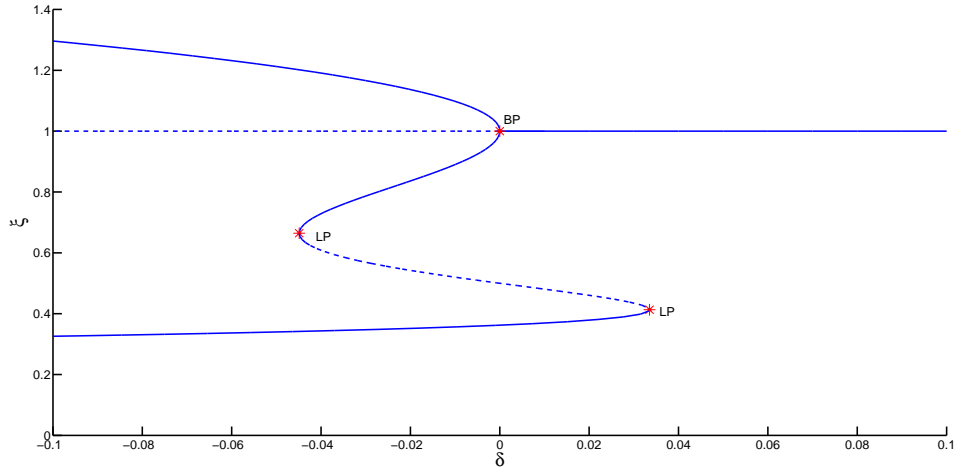}
  \caption{
    Numerical continuation of $\theta_0(\xi,a)=0$ from
    \eqref{genps5f1om} with the data from \eqref{n3exa1} and
    \eqref{n3exa2}. Pitchfork bifurcation at $(\delta_0,\xi_0)=(0,1)$
    (BP) and two saddle node bifurcations (LP) at
    $(\delta_-,\xi_-)=(-.04488...,.66691(4)...)$ and
    $(\delta_+,\xi_+)=(.03352...,.41262(522)...)$. 
    For $\delta=0$ one encounters the prescribed triple zero $\xi=1$,
    the zero $\xi=\frac{1}{2}$ and an additional zero near
    $.36222(562)...$. For $\delta=-.03$, one has 5 distinct $\xi$-values
    $\xi^{(j)}$ leading to 5 distinct steady states $b^{(j)}$ of
    \eqref{eq:ode_def} ($j=1,...,5$, cf. Table~\ref{numtab}). 
      Solid
      lines correspond to $\xi$'s yielding exponentially stable steady
      states, dashed lines to $\xi$'s yielding unstable steady states.
  }
  \label{pitch_5ss}
\end{figure}

A one-parameter continuation
\begin{equation}\nn 
a=a^*+\delta\, e_{10}\, , \quad -.05<\delta<.05\, ,
\end{equation}
in \eqref{genps5f1om} is leading to the bifurcation diagram in Figure~\ref{pitch_5ss} 
in the $(\delta,\xi)$-plane.

\begin{table}[h]
  \centering
  \begin{tabular}{|c|c|c|c|c|c|} \hline
    Phos. \#           & $b^{(1)}$ &$b^{(2)}$&$b^{(3)}$&$b^{(4)}\equiv a$&$b^{(5)}$ \\ \hline
    \multirow{3}{*}{0} & 1.4730 & 1.2198 & 1.0793 & 1 &  0.9618\\
                       & 4.7498 & 2.4000 & 1.4726 & 1 & 0.7700\\
                       & 4.2424 & 2.1440 & 1.3722 & 1 & 0.8246 \\ \hline
    \multirow{3}{*}{1} & 41.3012& 17.2813& 9.3826 & 5.9026 & 4.3718 \\
                       & 1.6493 & 1.3655 & 1.1583 & 1 & 0.8980\\
                       & 6.9970 & 2.9277 & 1.5895 & 1 &  0.7406\\ \hline
    \multirow{3}{*}{2} & 5.1859 & 3.5554 & 2.6688 & 2.1344 &   1.8438\\
                       & 0.5726 & 0.7768 & 0.9112 & 1 & 1.0474 \\
                       & 0.2429 & 0.1665 & 0.1250 & .1 & 0.0863 \\ \hline
    \multirow{3}{*}{3} & 209.9882& 235.8919 & 244.8175 & 248.9113 &  250.7710\\
                       & 0.0636  & 0.1414   & 0.2293   & .32 & 0.3909 \\
                       & 50.6175 & 56.8616  & 59.0132  & 60 & 60.4482 \\ \hline\hline
    \multirow{1}{*}{$\xi$}& 0.3472 & 0.5689 & 0.7866   & 1 & 1.1662 \\ \hline	       
  \end{tabular}
 
 \vspace{6mm} \caption{\label{numtab}The five admissible steady states  $b^{(j)}$
of \eqref{eq:ode_def} for $\delta=-.03$ and the corresponding zeros $\xi^{(j)}$
of \eqref{genps5f1om} up to 4 decimals: the numerical values of the 
rate constant vectors $\kappa=\kappa(a)$ and $\kappa(a^*)$ 
coincide up to the first 4 decimals, but the components 
$\kappa_{14}(a)=\kappa_{15}(a)= 0.00401749....$  and
$\kappa_{14}(a^*)=\kappa_{15}(a^*)= 0.00401701....$ differ.
}
\end{table}

For $\delta=-.03$, the numerical values for the five admissible zeros $\xi^{(j)}$
of \eqref{genps5f1om} and the five admissible steady states  $b^{(j)}$
of \eqref{eq:ode_def} can be found in Table~\ref{numtab}.

Numerical computations lead to the conclusion that $b^{(1)}$, $b^{(3)}$ and $b^{(5)}$
are exponentially stable steady states of \eqref{eq:ode_def} whereas
the Jacobian at $b^{(2)}$ as well as the Jacobian   at $b^{(4)}$
possesses one positive eigenvalue.

\begin{figure}[h]
  \centering
  \subfloat[Continuation for $-\,10^{-3}$ $\leq \delta \leq$ 
  $10^{-3}$ in the $(\delta,\xi)$-plane]{
    \includegraphics[width=0.49\linewidth]{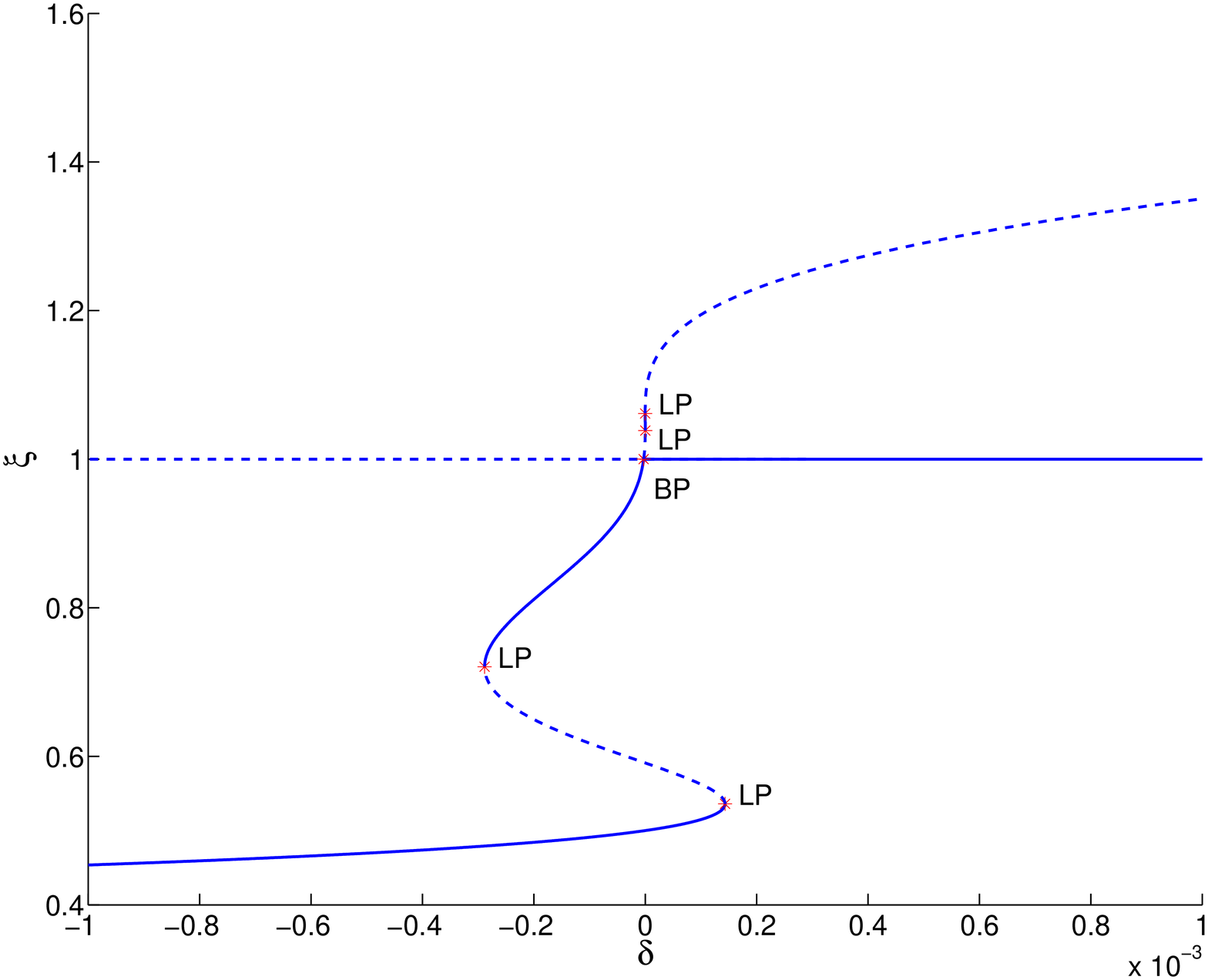}
  }
  \subfloat[Zoom to $-4 \cdot 10^{-6}$ $\leq \delta \leq$ $+4 \cdot 10^{-6}$; 
  cyan diamonds indicate six zeros for $\delta=0$]{
     \includegraphics[width=0.49\linewidth]{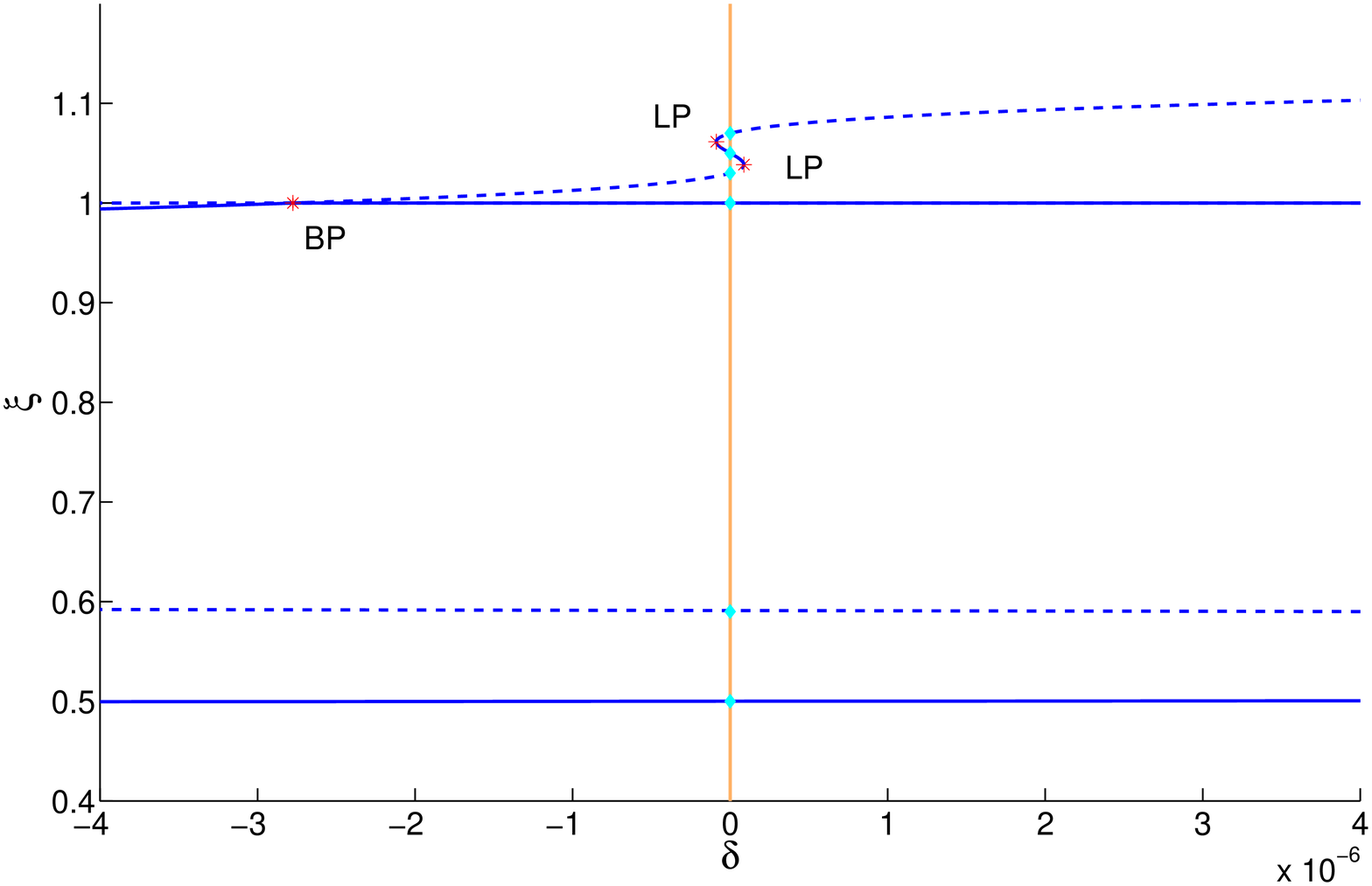}
  }
  \caption{
      Numerical continuation of $\theta_0(\xi,a)=0$ from
      \eqref{genps5f1om} with the data from
    \eqref{n4example} 
    showing 6 zeros $0.5$, $0.5910929...$, $1$, $1.03$, $1.05$ and
    $1.07$ 
     -- there is a 7th zero $51.07286...$ near $\xi=51$.
      Solid
      lines correspond to $\xi$'s yielding exponentially stable steady
      states, dashed lines to $\xi$'s yielding  unstable steady states.
      The label LP
      denotes  saddle-node bifurcation points, the label  BP 
      transcritical bifurcation points.
  }\label{nov28-7defs.1} 
\end{figure}

\vspace{3mm}For $n\geq 3$, the above argument can be applied to an 
$n$-site phosphorylation to create networks with $n+1$ steady states
for \eqref{eq:ode_def} by tuning the $n$ parameters $a_{3j+1}$, $j=1,2,...,n$.
For odd $n$, one is then, generically, expecting $n+2$ such steady states.
Using this rationale 
for even $n=4$, we have constructed a phosphorylation network
with a determining equation \eqref{genps5f1om} with 5 prescribed zeros
at $0.5$, $1$, $1.03$, $1.05$ and $1.07$ by choosing $a\in \Rp^{15}$
as
\eq{n4example}{llllll}{
a_1=1\, , &a_2=1\, , &a_3=1\, , &a_4=1.983448\, , &a_5=1\, , &a_6=1\, ,\\ 
a_7=469.6162955\, , &a_8=1\, , &a_9=400\, , &a_{10}=73.8036\, , &a_{11}=.32\, , &a_{12}=60\, ,\\
a_{13}=.5807998\, , &a_{14}=7\, , &a_{15}=1.8\, . &&&
}
As it turns out, this determining equation has two additional positive zeros,
one near $.59$ and one near $51.07$. See
Figure~\ref{nov28-7defs.1}. 

\vspace{0mm}
\section{\ The geometry of multistationarity}
\label{sec:discuss}
\renewcommand{\theequation}{\thesection.\arabic{equation}}
\setcounter{equation}{0}

  Here we discuss multistationarity and the constraints imposed
  on steady states within one coset of the
  stoichiometric subspace.

\subsection{
    Relation to sign patterns $s_1$, \ldots, $s_7$ from \cite{BMB2013}}\label{Diss1}\mbox{}
    
As a consequence of \cite{BMB2013}, any two distinct steady
  states $a$ and $b$ of (\ref{eq:ode_def}) (for $n$ arbitrary) within
  one  coset of the stoichiometric subspace satisfy the following:
  the sign pattern $\sign(\ln b/a)$ obeys one of the formulae $s_1$ -- $s_7$
  from \cite{BMB2013}. For the steady states of the
  3-site phosphorylation system we observe
that the sign vector for $ \ln\brac{b^{(j+1)}/b^{(j)}}$ is 
given by
$s_2:=(-,-,-|-,-,-|-,+,-|+,+,+)^T$
for $j=1,2,3,4$
so that these steady states are ordered with respect to $s_2$.

For the example with $n=4$ with steady states $b^{(j)}$ belonging to
increasing $\xi_j$ ($j=1,...,7$) with values $0.5, 0.59..., 1, 1.03,
1.05, 1.07, 51,...$ and $b^{(3)}=a$: the $\ln\brac{b^{(j)}/a}$,
$j=1,2$, belong to  $-s_1$, the $\ln\brac{b^{(j)}/a}$, $j=4,5,6$,
belong to  $s_1:=(+,-,+|-,-,-|-,-,-|+,+,+|+,+,+)$.
Finally, $\ln\brac{b^{(7)}/a}$ belongs to
$s_5:=(+,-,+|-,-,-|-,-,-|+,-,+|+,+,+)$.\\ 
Moreover, the
$\ln\brac{b^{(j+1)}/b^{(j)}}$ belong to the sign patterns  
$s_7:=(+,-,-|-,-,-|-,+,-|+,+,+|+,+,+)$ for $j=1,5$, to $s_1$ for 
$j=2,3,4$ and to $s_5$ for $j=6$. 

\subsection{
Geometric constraints on multistationarity    
  }\label{Diss2}
\mbox{}

According to the
  ordering of variables in (\ref{tab:VarAssignment}), we introduce the
  following  notation for $g^L=\frac{b}{a}$ with the matrix $L$ from (\ref{eq:def_Ln}):
  \begin{equation}\nn 
    g^L=\Big(
    \Gamma_{E_1},\Gamma_{A},\Gamma_{E_2}|\,
    \Gamma_{AE_1},\Gamma_{A_P},\Gamma_{A_pE_2}|\,
    \Gamma_{A_PE_1},\Gamma_{A_{2P}},\Gamma_{A_{2P}E_2}|\ \ldots \ |\,
    \Gamma_{A_{(n-1)P}E_1},\Gamma_{A_{nP}},\Gamma_{A_{nP}E_2}\Big)^T
  \end{equation}
  with 
  \begin{align*} 
    \Gamma_{E_1} &= (g^L)_1 \,=\, \frac{g_1g_2^{n-1}}{g_3}\, , & 
    \Gamma_A &= (g^L)_2 \,=\, \frac{1}{g_1g_2^n}\, ,       &
    \Gamma_{E_2} &= (g^L)_3 \,=\, \frac{g_1g_2^{n-2}}{g_3}\, ,\\
    \Gamma_{A_{i-1P}E_1} &= (g^L)_{1+3i} \,=\, \frac{g_2^i}{g_2^2g_3}\, , & 
    \Gamma_{A_{iP}} &= (g^L)_{2+3i} \,=\, \frac{g_2^i}{g_1g_2^n}\, ,  &
    \Gamma_{A_{iP} E_2} &= (g^L)_{3+3i} \,=\, (g^L)_{1+3i}
  \end{align*}
  for $i=1,...,n$. We recall the form
   \begin{displaymath}\label{genps4a+}
    g_1 = \xi^{1-n}F_1(\xi)/\Delta(\xi),\ g_2\equiv \xi\,,\
    g_3 = \xi^{-1}F_3(\xi)/\Delta(\xi)
  \end{displaymath}
  of the \eqref{genps3}-solutions
  where  $\xi$ is to be a positive zero of (\ref{genps5f1}) or 
  (\ref{genps5f1om}) (cf.~\eqref{genps4}).  
  So we obtain for the partitioning 
  \begin{equation}\label{biogl}
    g^L=\left(\left.\Gamma_{(0)}^T\right|\,
      \left.\Gamma_{(1)}^T\right|\,
      \left.\phantom{\frac{.}{.}}\ldots\,\phantom{\frac{.}{.}}\right|
     \Gamma_{(n)}^T\right)^T\in\Rp^{3+3n}
  \end{equation}
the following identities: 
  \begin{equation}\label{biogli0}
    \Gamma_{(0)}^T:=\big(\Gamma_{E_1},\,\Gamma_{A},\,\Gamma_{E_2}\big)\,= \,
    \Big(\xi\, \frac{F_1(\xi)}{F_3(\xi)},\,
    \frac{\Delta(\xi)}{\xi\, F_1(\xi)},\,
    \frac{F_1(\xi)}{F_3(\xi)}\Big)\,,\quad
    \xi=\dfrac{\Gamma_{E_1}}{\Gamma_{E_2}}\, , 
  \end{equation}
   \begin{equation}\label{biogli01}
    \Gamma_{(1)}^T:=\big(\Gamma_{AE_1},\,\Gamma_{A_P},\,\Gamma_{A_PE_2}\big)\,= \,
    \Big(\frac{\Delta(\xi)}{F_3(\xi)},\,
    \frac{\Delta(\xi)}{\xi\, F_1(\xi)},\,
    \frac{\Delta(\xi)}{F_3(\xi)}\Big)=
    \Big(\Gamma_{A}\Gamma_{E_1},\xi\Gamma_{A},\Gamma_{A_P}\Gamma_{E_2}\Big)\,.
   \end{equation} 
   
   \vspace*{1mm}
\begin{equation}\label{biogli0i}   
   \Gamma_{(i)}^T:=
   \big(\Gamma_{A_{(i-1)P}E_1},\Gamma_{A_{iP}},\Gamma_{A_{iP}E_2}\big)=
   \xi^{i-1}\big(\Gamma_{AE_1},\,\Gamma_{A_P},\,\Gamma_{A_PE_2}\big)=
   \xi^{i-1}\Gamma_{(1)}^T
    \end{equation}
\begin{subequations}\label{homogenuity}
In particular one has for $i=1,\ldots, n$:
  \begin{eqnarray}
    \label{eq:kin_comp}
    \Gamma_{A_{i-1 P} E_1} &=& \xi^{i-1} \Gamma_A \Gamma_{E_1} \, ,\\\label{eq:phs_comp}
    \Gamma_{A_{i P}} &= & \xi^{i-1} \Gamma_{A_P}\phantom{\Gamma_{E_2}}\ = \xi^{i}\Gamma_{A}\, ,\\\label{eq:phos_form}
    \Gamma_{A_{i P} E_2} &=& \xi^{i-1} \Gamma_{A_P}\Gamma_{E_2}\ = \xi^{i} 
    \Gamma_{A}\Gamma_{E_2}= \xi^{i-1} \Gamma_A \Gamma_{E_1}
  \end{eqnarray}
  \end{subequations}

We summarize these geometric properties in the following fact:
\begin{fact}\label{fact:computing_Gamma}
  Let $\kappa\in\Rp^{6n}$ be given and assume
  network~(\ref{eq:network_dd}) admits multistationarity, that is,
  there exists  two distinct positive vectors $a$ and $b$ such that
  \begin{displaymath}
    S\, r(\kappa,a)=S\, r(\kappa,b)=0,\; Z\, (b-a)=0.
  \end{displaymath}
  Then the steady state concentrations $a_1$ and $b_1$ of the
  kinase together with the steady state concentrations $a_3$ and
  $b_3$ of the phosphatase and $a_2$ and $b_2$ of the
  unphosphorylated protein allow the reconstruction of the ratios
  $(g^L)_i=\frac{b_i}{a_i}$, $i=4$, \ldots, $3+3n$, in the following way:
  \begin{displaymath}
    \Gamma_{(0)}^T=\big(\Gamma_{E_1}, \Gamma_A,\Gamma_{E_2}\big)=
    \Big(\frac{b_1}{a_1},\,\frac{b_2}{a_2},\,\frac{b_3}{a_3}\Big)\quad \text{and}\quad
    \xi=\frac{\Gamma_{E_1}}{\Gamma_{E_2}}\,  = \, \dfrac{b_1/a_1}{b_2/a_3}\, ,
  \end{displaymath}
  with 
  \begin{displaymath}\Gamma_{(1)}^T
    =\big(\Gamma_{A}\Gamma_{E_1},\,\xi\Gamma_{A},\,\xi\Gamma_{A}\Gamma_{E_2}\big)
    =\Big(\frac{b_4}{a_4},\,\frac{b_5}{a_5},\,\frac{b_6}{a_6}\Big)
  \end{displaymath}
  and 
  \begin{displaymath} 
    \Gamma_{(i)}^T=
    \big(\Gamma_{A_{(i-1)P}E_1},\Gamma_{A_{iP}},\Gamma_{A_{iP}E_2}\big)
    = \xi^{i-1} \Big( \frac{b_4}{a_4},\, \frac{b_5}{a_5},\,
    \frac{b_6}{a_6} \Big) = \Big( \frac{b_{1+3i}}{a_{1+3i}},\,
    \frac{b_{2+3i}}{a_{2+3i}},\, \frac{b_{3+3i}}{a_{3+3i}} \Big)
  \end{displaymath} 
  for $i=1,...,n$. In particular one has for $i=1,...,n-1$ 
  \begin{equation}\label{biogli4}
    \xi=\dfrac{\Gamma_{E_1}}{\Gamma_{E_2}}=
      \dfrac{\Gamma_{A_P}}{\Gamma_{A}}=
      \dfrac{\Gamma_{A_{(i+1)P}}}{\Gamma_{A_{iP}}}=
      \dfrac{\Gamma_{A_{iP}E_1}}{\Gamma_{A_{(i-1)P}E_1}}=
      \dfrac{\Gamma_{A_{(i+1)P}E_2}}{\Gamma_{A_{iP}E_2}}\, .
  \end{equation}
\end{fact}

\subsection{
  Reconstruction of steady state ratios from measured
  kinase $E_1$, phosphatase $E_2$ and substrate $A$
    }\label{Diss3}
\mbox{}

Consider the experimental investigation of a specific multisite
phosphorylation system~(\ref{eq:network_dd}) whereby the rate
constants $\kappa$ and the total concentrations are fixed, but might
not (all) be known. Suppose we know a priory that the system exhibits 
multistationarity for the given rate constants and total
concentrations. Then steady state data of the concentration of kinase,
phosphatase and protein in two different steady states $a$ and $b$
(for these total concentrations) are sufficient to reconstruct all
fractions $\frac{b_i}{a_i}$ of the two steady states. That is, it
suffices to measure $a_1$, $a_2$, $a_3$ and $b_1$, $b_2$, $b_3$ to
reconstruct all the ratios $\frac{b_i}{a_i}$, $i=1$, \ldots, $3+3n$.

\subsection{A graphical test for the coset condition}
\label{Diss4}
\mbox{}

Next we elaborate on (\ref{biogli4}). For the steady state
concentrations of the phosphoforms $a_{3i+2}$ and $b_{3i+2}$, it
implies  
\begin{displaymath}
  \frac{b_{3i+2}}{b_{3i-1}} = \xi\ \frac{a_{3i+2}}{a_{3i-1}}\ \text{ for  $i=1$, \ldots, $n$.}
\end{displaymath}
Hence the fractions $\frac{b_{3i+2}}{b_{3i-1}}$ and
$\frac{a_{3i+2}}{a_{3i-1}}$ are collinear. Likewise we find for the
fractions of kinase substrate  and of phosphatase substrate
complexes
\begin{displaymath}
  \frac{b_{3i+1}}{b_{3i-2}} = \xi\ \frac{a_{3i+1}}{a_{3i-2}} \
  \text{ and }\ \frac{b_{3i+3}}{b_{3i}} = \xi\
  \frac{a_{3i+3}}{a_{3i}}\ \text{ for  $i=1$, \ldots, $n$.}
\end{displaymath}
We summarize this in the following fact:
\begin{fact}[Collinearity of {\em relative} steady
  states]\label{fact:collinear} \mbox{}\\
  Given $\kappa\in\Rp^{6n}$ and steady states $a$, $b\in\Rp^{3+3n}$ of
  (\ref{eq:ode_def}), we define 
  \begin{displaymath}
    \alpha_i:= \frac{a_{i+3}}{a_{i}}\, \quad \beta_i:=
    \frac{\Red{b_{i+3}}}{b_{i}}\, ,\quad i=1,\, \ldots,\, 3n\, .
  \end{displaymath}
  If $a$ and $b$ belong to the same coset (i.e., $Z\, (b-a)=0$),
  then the pairs ($\alpha_i$, $\beta_i$) are collinear, i.e., the
  pairs ($\alpha_i$, $\beta_i$) are on the line $\beta=\xi\,\alpha$
  with slope $\xi = \frac{b_1/a_1}{b_3/a_3}$.
\end{fact}

\begin{remark}[Graphical test for steady states to satisfy the coset
  condition]\label{rem:graph_excl} \mbox{}\\ \rm
  Suppose for the phosphoforms $A$, $A_P$, \ldots, $A_{nP}$ two
  different sets of  steady state values have been measured (i.e.,
  there exists data for $a_2$, $a_5$, \ldots, $a_{2+3n}$ and $b_2$,
  $b_5$, \ldots, $b_{2+3n}$). If these belong to two steady states
  within one and the same coset (i.e., are components of two steady
  states $a$, $b\in\Rp^{3+3n}$ with $Z\, (b-a)=0$), then the points
  \begin{displaymath}
    \alpha_i:= \frac{a_{3i+2}}{a_{3i-1}}\, ,\quad \beta_i:=
    \frac{b_{3i+2}}{b_{3i-1}}\, ,\quad  i=1,\, \ldots,\, n\, , 
  \end{displaymath}
  are collinear. Hence, when one measures  two steady state values of
  $A$, \ldots, $A_{nP}$ so that the points ($\alpha_i$,$\beta_i$) are
  not collinear then these two steady states do not give rise to 
  multistationa\-rity.
\end{remark}

\bibliographystyle{plain}
\bibliography{5ss}

\vspace{-0mm}

\appendix

\section{\ The network matrices for $n\geq 2$}
\label{sec:smat}

The matrices $\Y$, $Z$, $E$ and $L$ can be
  obtained from eqs.~(\ref{eq:Y_direct}), (\ref{eq:def_Zi}),  (\ref{eq:Ei}) and
  (\ref{eq:def_Ln}) of this manuscript.
  We recall the definition of  the stoichiometric matrix $S$ from Section~3 
  of \cite{BMB2013}.
  With the following sub-matrices
  \begin{align}\nn
    n_{11} &= \left[
      \begin{array}{rrrrrr}
        -1& \phantom{-}1 & \phantom{-}1 & \phantom{-}0 & \phantom{-}0 &
        \phantom{-}0 \\ 
        -1& 1 & 0 & 0 & 0 & 1 \\
        0 & 0 & 0 & -1& 1 & 1 
      \end{array}
    \right],\ \
    &n_{12} &= \left[
      \begin{array}{rrrrrr}
        -1 & \phantom{-}1 & \phantom{-}1 & \phantom{-}0 & \phantom{-}0 &
        \phantom{-}0   \\ 
        0 & 0 & 0 & 0 & 0 & 0   \\
        0 & 0 & 0 & -1& 1 & 1 
      \end{array}
    \right],\\\nn
    n_{21} &= \left[
      \begin{array}{rrrrrr} 
        \phantom{-}1 & -1& -1& \phantom{-}0 & \phantom{-}0 & \phantom{-}0 \\
        0 & 0 & 1 & -1& 1 & 0 \\
        0 & 0 & 0 & 1 & -1& -1 
      \end{array}
    \right],\ \
    & n_{22} &= \left[
      \begin{array}{rrrrrr}
        \phantom{-}0 & \phantom{-}0 & \phantom{-}0 & \phantom{-}0 &
        \phantom{-}0 & \phantom{-}0 \\ 
        -1& 1 & 0 & 0 & 0 & 1 \\
        0 & 0 & 0 & 0 & 0 & 0 
      \end{array}
    \right].
  \end{align}
  of dimension $3\times 6$, one has 
  \begin{equation}\nn 
    S:= \ \left[ 
      \begin{array}{c|c|c|c|c|c}
        n_{11} & n_{12} & n_{12} & n_{12} & & n_{12} \\
        n_{21} & n_{22} & 0_{3\times 6} & \multirow{2}{*}{$0_{2\cdot
            3\times 6}$} & &
        \multirow{5}{*}{$0_{\brac{n-2}\cdot 3\times 6}$} \\
        \cline{1-1}
        0_{3\times 6} & n_{21} & n_{22} & \phantom{0_{3\times 6}} &
        \dots & \\ \cline{1-2}
        \multicolumn{2}{c|}{0_{3\times 6\cdot 2}} & n_{21} & n_{22}
        & & \\ \cline{1-3}
        \multicolumn{3}{c|}{0_{3\times 6\cdot 3}} & n_{21} & & \\ 
        \cline{1-4}
        \multicolumn{4}{c|}{\vdots} &\ddots & \\ \cline{1-5}
        \multicolumn{5}{c|}{0_{3\times 6\cdot \brac{n-1}}} & n_{21}
      \end{array}
    \right]\ \in \R^{\brac{3+3n} \times 6n}\, .
  \end{equation}

\vspace{3mm}
For the convenience of the reader, we
close this appendix with the data for $n=3$:

{\small
\begin{equation}\nn
S=   \left[
     \begin{array}{rrr|rrr|rrr|rrr|rrr|rrr}
-1&	1&	1&	0&	0&	0&	-1&	1&	1&	0&	0&	0&	-1&	1&	1&	0&	0&	0\\
-1&	1&	0&	0&	0&	1&	0&	0&	0&	0&	0&	0&	0&	0&	0&	0&	0&	0\\
0&	0&	0&	-1&	1&	1&	0&	0&	0&	-1&	1&	1&	0&	0&	0&	-1&	1&	1\\\hline
1&	-1&	-1&	0&	0&	0&	0&	0&	0&	0&	0&	0&	0&	0&	0&	0&	0&	0\\
0&	0&	1&	-1&	1&	0&	-1&	1&	0&	0&	0&	1&	0&	0&	0&	0&	0&	0\\
0&	0&	0&	1&	-1&	-1&	0&	0&	0&	0&	0&	0&	0&	0&	0&	0&	0&	0\\\hline
0&	0&	0&	0&	0&	0&	1&	-1&	-1&	0&	0&	0&	0&	0&	0&	0&	0&	0\\
0&	0&	0&	0&	0&	0&	0&	0&	1&	-1&	1&	0&	-1&	1&	0&	0&	0&	1\\
0&	0&	0&	0&	0&	0&	0&	0&	0&	1&	-1&	-1&	0&	0&	0&	0&	0&	0\\\hline
0&	0&	0&	0&	0&	0&	0&	0&	0&	0&	0&	0&	1&	-1&	-1&	0&	0&	0\\
0&	0&	0&	0&	0&	0&	0&	0&	0&	0&	0&	0&	0&	0&	1&	-1&	1&	0\\
0&	0&	0&	0&	0&	0&	0&	0&	0&	0&	0&	0&	0&	0&	0&	1&	-1&	-1
\end{array}\right]\, ,\end{equation}

\vspace{3mm}
\begin{equation}\nn
\Y^T= \left[
      \begin{array}{rrr|rrr|rrr|rrr}
1&1&0&0&0&0&0&0&0&0&0&0\\
0&0&0&1&0&0&0&0&0&0&0&0\\
0&0&0&1&0&0&0&0&0&0&0&0\\\hline
0&0&1&0&1&0&0&0&0&0&0&0\\
0&0&0&0&0&1&0&0&0&0&0&0\\
0&0&0&0&0&1&0&0&0&0&0&0\\\hline
1&0&0&0&1&0&0&0&0&0&0&0\\
0&0&0&0&0&0&1&0&0&0&0&0\\
0&0&0&0&0&0&1&0&0&0&0&0\\\hline
0&0&1&0&0&0&0&1&0&0&0&0\\
0&0&0&0&0&0&0&0&1&0&0&0\\
0&0&0&0&0&0&0&0&1&0&0&0\\\hline
1&0&0&0&0&0&0&1&0&0&0&0\\
0&0&0&0&0&0&0&0&0&1&0&0\\
0&0&0&0&0&0&0&0&0&1&0&0\\\hline
0&0&1&0&0&0&0&0&0&0&1&0\\
0&0&0&0&0&0&0&0&0&0&0&1\\
0&0&0&0&0&0&0&0&0&0&0&1
\end{array}\right] \, ,     
\end{equation}

\vspace{3mm}
\begin{equation}\nn 
E=\left[
        \begin{array}{rrr|rrr|rrr}
1&	0&	1&	0&	0&	0&	0&	0&	0\\
1&	0&	0&	0&	0&	0&	0&	0&	0\\
0&	0&	1&	0&	0&	0&	0&	0&	0\\\hline
0&	1&	1&	0&	0&	0&	0&	0&	0\\
0&	1&	0&	0&	0&	0&	0&	0&	0\\
0&	0&	1&	0&	0&	0&	0&	0&	0\\\hline
0&	0&	0&	1&	0&	1&	0&	0&	0\\
0&	0&	0&	1&	0&	0&	0&	0&	0\\
0&	0&	0&	0&	0&	1&	0&	0&	0\\\hline
0&	0&	0&	0&	1&	1&	0&	0&	0\\
0&	0&	0&	0&	1&	0&	0&	0&	0\\
0&	0&	0&	0&	0&	1&	0&	0&	0\\\hline
0&	0&	0&	0&	0&	0&	1&	0&	1\\
0&	0&	0&	0&	0&	0&	1&	0&	0\\
0&	0&	0&	0&	0&	0&	0&	0&	1\\\hline
0&	0&	0&	0&	0&	0&	0&	1&	1\\
0&	0&	0&	0&	0&	0&	0&	1&	0\\
0&	0&	0&	0&	0&	0&	0&	0&	1\\
\end{array}
      \right]\quad \mbox{with} \quad SE=0\, ,\end{equation}
      
\vspace{3mm}
\begin{equation}\nn
L= (L_1,L_2,L_3)=\left[
      \begin{array}{r|rr}
1& 2&-1\\
-1&-3&0\\
1&1&-1\\\hline 
0&-1&-1\\
-1&-2&0\\    
0&-1&-1\\\hline 
0&0&-1\\
-1&-1&0\\    
0&0&-1\\\hline 
0&1&-1\\
-1&0&0\\    
0&1&-1
\end{array}\right] \quad \mbox{with} \quad \Y^TL_1=0 \ .
\end{equation}
}

\end{document}